\documentstyle[aps]{revtex}

\begin{document}
\author{E. J. Ferrer$^{1},$ V. de la Incera$^{1},$ and A. Romeo$^{2}$}
\address{$^{1}$Department of Physics, State University of New York at Fredonia,\\
Houghton Hall 118, Fredonia, NY 14063, USA\\
$^{2}$Institute for Space Studies of Catalonia, CSIC, Edif. Nexus, Gran\\
Capita 2-4, 08034 Barcelona, Spain.}
\title{Photon Propagation in Space-Time with a Compactified Spatial Dimension}
\maketitle

\begin{abstract}
The one-loop effects of vacuum polarization induced by untwisted fermions in
QED in a nonsimply connected space-time with topology $S^{1}\times R^{3}$
are investigated. It is found that photon propagation in this system is
anisotropic, appearing several massive photon modes and a superluminal
transverse mode. For small compactification radius $a$, the superluminal
velocity increases logarithmically with $a$. At low energies the photon
masses lead to an effective confinement of the gauge fields into a
(2+1)-dimensional manifold transverse to the compactified direction. The
system shows a topologically induced directional superconductivity.
\end{abstract}

It is well known that the global properties of the space-time,
even if it is locally flat, can give rise to new physics. The
seminal discovery in this direction is the so called Casimir
effect \cite{Casimir}$,$\cite{Milton}.

A characteristic of quantum field theory in space-time with non-trivial
topology is the existence of non-equivalent types of fields with the same
spin\cite{Isham}$.$ In particular, for a fermion system in space-time which
is locally flat but with $S^{1}\times R^{3}$ topology (i.e. a Minkowskian
space with one of the spatial dimensions compactified in a circle $S^{1}$ of
finite radius $a$), the nontrivial topology is transferred into periodic
boundary conditions for untwisted fermions or antiperiodic boundary
conditions for twisted fermions\cite{Ford}

\begin{equation}
\psi \left( t,x,y,z\right) =\pm \psi \left( t,x,y,z+a\right)  \label{1}
\end{equation}

Due to polarization effects of untwisted electrons, quantum electrodynamics
(QED) in $S^{1}\times R^{3}$ space-time with photons coupled to untwisted
fermions or to a combination of twisted and untwisted fermions is known to
have an unstable zero vacuum\cite{Ford}$,$\cite{Ford2}. It can be shown,
however, that a constant expectation value of the electromagnetic potential
along the compactified direction

\begin{equation}
\overline{A}_{\nu }=\Lambda \delta _{\nu 3},\qquad \Lambda =\frac{\pi }{ea}%
\left(
\mathop{\rm mod}%
2\pi \right)  \label{2}
\end{equation}
minimizes the effective potential stabilizing the theory. The same stable
vacuum solution was found in Ref. \cite{Hosotani} in the case of massless
QED with periodic fermions on a circle (QED with $S^{1}\times R^{1}$
topology). Notice that even though such a vacuum configuration has $F_{\mu
\nu }=0$, it cannot be gauged to zero, because the gauge transformation that
would be needed does not respect the periodicity of the function space in
the $S^{1}\times R^{3}$ domain. This is a sort of Aharonov-Bohm effect which
makes $A_{\mu }$ a dynamical variable due to the non-simply connected
topology of the considered space-time. The non-gauge equivalence between a
constant component of the gauge potential ($A_{0}$ in this case) and zero is
also present in QED at finite temperature and/or density, due to the
compactification of the time coordinate\cite{Batakis}. The minimum of the
potential in the statistical case is however at $A_{0}=0$, since only
twisted fermions are allowed in statistics. On the other hand, in the
electroweak theory with a finite density of fermions, a non-trivial constant
vacuum $A_{0}$ is induced by the fermion density and cannot be gauged away
\cite{Linde}. There, in contrast to the system considered in the present
paper, an additional parameter (a leptonic and/or baryonic chemical
potential) is needed to trigger the non-trivial constant minimum for $A_{0}$.

Our main goal in the present work is to investigate the photon propagation
in the $S^{1}\times R^{3}$ domain. We expect that the non-simply connected
character of the space-time may give rise to new physics which would
normally be absent in the flat space with trivial topology. To understand
this in qualitative terms one should have in mind that, thanks to the vacuum
polarization, the photon exists during part of the time as a virtual $%
e^{+}e^{-}-pair$. The virtual pairs can then transfer to the real photon the
properties of the quantum vacuum which, as known, depend on the non-trivial
topology and boundary conditions of the space under consideration. Thus, to
study the photon propagation one needs to investigate the effects connected
to vacuum polarization in the $S^{1}\times R^{3}$ space-time.

The vacuum polarization in the non-trivial spatial topology $S^{1}\times
R^{3}$ can be influenced by both virtual untwisted and twisted $%
e^{+}e^{-}-pairs$. The results for twisted fermions can be easily read off
the results at finite temperature, since in the Euclidean space the two
theories are basically the same after the interchange of the four-space
subindexes $3\leftrightarrow 4$. Nevertheless, a different situation occurs
with untwisted fermions that has no analogy in the statistical case.
Henceforth, we concentrate our attention in the untwisted-fermion case.

As shown in this paper, when vacuum polarization due to untwisted fermions
is considered, various new physical effects emerge. First, at very small
radius of compactification, the photon propagation is effectively confined
to a Minkowskian $(2+1)-$dimensional manifold on which only superluminal
photons propagate. This means that the photons moving in such a lower
dimensional space can feel the lack of Lorentz symmetry of the general
manifold ($S^{1}\times R^{3}$) on which the lower dimensional space is
embedded, so it is allowed for them to have group velocity larger than the
usual Minkowskian velocity $c$. Another novel feature of our results is that
the photon shows an anisotropic magnetic response. This anisotropy gives
rise to a directional superconductivity, with the supercurrent flowing in
the direction of the compactified direction. The magnetic penetration length
has a purely topological character (i.e. it is a function only of the
compactification length $a$). The results here reported can be of interest
for theories with extra dimensions on which some of them are compactified%
\cite{ExtraD}, and for condensed matter quasi-planar systems\cite{ConMat}.

To study the photon propagation we should solve the dispersion relations of
the electromagnetic modes, which in the low-frequency limit has the general
form

\begin{equation}
k_{0}^{2}-{\bf k}^{2}+\Pi ({\bf k}^{2})=0  \label{2a}
\end{equation}
where $\Pi ({\bf k}^{2})$ accounts for vacuum polarization effects.
Different external conditions, as external fields, geometric boundary
conditions, thermal bath, etc., can modify the vacuum and produce, through $%
\Pi ({\bf k}^{2})$, a variation in the spectrum of the photon modes.

The solution of the photon dispersion equations (\ref{2a}) can be obtained
as the poles of the photon Green's function. Due to the explicit Lorentz
symmetry breaking in the $S^{1}\times R^{3}$ topology, in addition to the
usual tensor structures $k_{\mu }$ and $g_{\mu \nu }$, a spacelike unit
vector pointing along the compactified direction $n^{\mu }=(0,0,0,1)$ must
be considered. Then, the general structure of the electromagnetic field
Green's function is

\begin{equation}
\Delta _{\mu \nu }(k)=P(g_{\mu \nu }-\frac{k_{\mu }k_{\nu }}{k^{2}})+Q[\frac{%
k_{\mu }k_{\nu }}{k^{2}}-\frac{k_{\mu }n_{\nu }+n_{\mu }k_{\nu }}{(k\cdot n)}%
+\frac{k^{2}n_{\mu }n_{\nu }}{(k\cdot n)^{2}}]+\frac{\alpha }{k^{4}}k_{\mu
}k_{\nu },  \label{8}
\end{equation}
where $\alpha $ is a gauge fixing parameter corresponding to the covariant
gauge condition $\frac{1}{\alpha }\partial _{\mu }A_{\mu }=0$, and $P$ and $%
Q $ are defined as

\begin{equation}
P=\frac{1}{k^{2}+\Pi _{0}},\qquad Q=-\frac{\Pi _{1}}{(k^{2}+\Pi _{0})\left\{
k^{2}+\Pi _{0}-\Pi _{1}[k^{2}/(k\cdot n)^{2}+1]\right\} },  \label{9}
\end{equation}
The parameters $\Pi _{0}$ and $\Pi _{1}$ are the coefficients of the
polarization operator $\Pi _{\mu \nu }$, which in the $S^{1}\times R^{3}$
space can be written as

\begin{equation}
\Pi _{\mu \nu }(k)=\Pi _{0}(g_{\mu \nu }-\frac{k_{\mu }k_{\nu }}{k^{2}})+\Pi
_{1}[\frac{k_{\mu }k_{\nu }}{k^{2}}-\frac{k_{\mu }n_{\nu }+n_{\mu }k_{\nu }}{%
(k\cdot n)}+\frac{k^{2}n_{\mu }n_{\nu }}{(k\cdot n)^{2}}]  \label{10}
\end{equation}

From (\ref{8})-(\ref{9}) the photon dispersion relations are

\begin{equation}
k_{0}^{2}-{\bf k}^{2}+\Pi _{0}=0  \label{11}
\end{equation}

\begin{equation}
k_{0}^{2}-{\bf k}^{2}+(\Pi _{0}-\frac{\widehat{k}^{2}}{k_{3}^{2}}\Pi _{1})=0
\label{12}
\end{equation}
where $\widehat{k}^{2}=k_{0}^{2}-k_{\bot }^{2}$ and $k_{\bot
}^{2}=k_{1}^{2}+k_{2}^{2}$. Notice that in addition to the transverse mode
associated to Eq. (\ref{11}) (normally present in Minkowski space-time with
trivial topology), a longitudinal mode, Eq. (\ref{12}), arises here due to
the presence of the extra coefficient $\Pi _{1}$. The situation resembles
the finite temperature case. Nevertheless, as discussed below, the physical
consequences of the spatial compactification are radically different from
those already known at finite temperature.

Since the compactified ${\cal OZ}$-direction distinguishes itself from the
other spatial directions, we should separate the analysis for photons
propagating along ${\cal OZ}$ ($k_{\perp }=0$), from those propagating
perpendicularly to it ($k_{3}=0$).

The dispersion relations (\ref{11}) and (\ref{12}) for photons propagating
perpendicularly to the compactified direction ($k_{3}=0$) are found, from
Eqs.(\ref{10})-(\ref{12}), to reduce respectively to

\begin{equation}
k_{0}^{2}-k_{\perp }^{2}-\frac{\widehat{k}^{2}}{k_{\bot }^{2}}\Pi _{00}=0
\label{13}
\end{equation}

\begin{equation}
k_{0}^{2}-k_{\perp }^{2}-\Pi _{33}=0  \label{14}
\end{equation}

Let us find the solutions of Eqs. (\ref{13})-(\ref{14}) at the one-loop
level. With this end, we need to calculate the one-loop polarization
operator components $\Pi _{00}$ and $\Pi _{33}$ for untwisted fermions.
Considering the free propagator of untwisted fermions on the minimum
solution (\ref{2})

\begin{equation}
\widetilde{G}(x-x^{\prime })=\frac{1}{(2\pi )^{3}a}%
\sum\hspace{-0.5cm}\int%
d^{4}p\exp [ip\cdot (x-x^{\prime })]G(\widetilde{p})  \label{14a}
\end{equation}
where

\begin{equation}
G(\widetilde{p})=\frac{\widetilde{p}\llap / -m}{\widetilde{p}%
^{2}-m^{2}+i\epsilon },\qquad \widetilde{p}_{\mu }=\left( p_{0},p_{\perp
},p_{3}-e\Lambda \right)  \label{4}
\end{equation}
and $%
\sum\hspace{-0.4cm}\int%
d^{4}p=\sum\limits_{p_{3}}\int d^{3}p,$ $p_{3}=2n\pi /a,$ $(n=0,\pm 1,\pm
2,...)$ being the discrete frequencies associated to periodic fermions, the
corresponding one-loop polarization operator is given by

\begin{equation}
\Pi _{\mu \nu }(k)=-\frac{4ie^{2}}{(2\pi )^{3}a}%
\sum\hspace{-0.5cm}\int%
d^{4}p\frac{\widetilde{p}_{\mu }(\widetilde{p}_{\nu }-k_{\nu })+\widetilde{p}%
_{\nu }(\widetilde{p}_{\mu }-k_{\mu })-\widetilde{p}\cdot (\widetilde{p}%
-k)g_{\mu \nu }+m^{2}g_{\mu \nu }}{(\widetilde{p}^{2}-m^{2})\left[ (%
\widetilde{p}-k)^{2}-m^{2}\right] }.  \label{15}
\end{equation}

In the $a\left| \widehat{k}\right| \ll am\ll 1$ limit, we obtain

\begin{equation}
\Pi _{00}(k_{3}=0,k_{0}=0,k_{\bot }\sim 0)\simeq \frac{e^{2}}{3\pi ^{2}}%
k_{\bot }^{2}\left[ \frac{1}{2}\ln \xi +{\cal O}(\xi ^{0})\right] +{\cal O}%
(k_{\bot }^{4})  \label{28}
\end{equation}
\begin{equation}
\Pi _{33}(k_{3}=0,k_{0}=0,k_{\bot }\sim 0)\simeq \frac{e^{2}}{a^{2}}\left[
\frac{1}{3}+{\cal O}(\xi ^{2})\right] +{\cal O}(k_{\bot }^{2})  \label{28a}
\end{equation}
where $\xi =am/2\pi \ll 1$ . Using the results (\ref{28}) and (\ref{28a}) in
the dispersion equations (\ref{13}), (\ref{14}), and taking into account
that the photon velocity for each propagation mode can be obtained from $%
{\rm v}({\bf k})=\partial k_{0}/\partial \left| {\bf k}\right| $ \footnote{%
In the considered low-frequency approximation the group and phase velocities
coincide and can be found by the same proposed formula.}, we find that
within the considered approximation the transverse and longitudinal modes
propagate perpendicularly to the compactified direction with velocities

\begin{equation}
{\rm v}_{\bot }^{T}\simeq 1-\frac{e^{2}}{12\pi ^{2}}\ln \xi ,  \label{18}
\end{equation}
\begin{equation}
{\rm v}_{\bot }^{L}\simeq 1-\left[ (M_{\bot }^{L})^{2}/2k_{\bot }^{2}\right]
,  \label{18a}
\end{equation}
respectively, where $(M_{\bot }^{L})^{2}=\Pi _{33}=e^{2}/3a^{2}>0$ plays the
role of an effective topological mass for the longitudinal mode.

It is worth to mention that he modifications found for the two velocities, $%
{\rm v}_{\bot }^{T}$ and ${\rm v}_{\bot }^{L}$, have different origins. The
modification of the longitudinal velocity ${\rm v}_{\bot }^{L}$ is due to
the appearance of the topological mass $M_{\bot }^{L}$; while the transverse
superluminal velocity ${\rm v}_{\bot }^{T}$ (note that ${\rm v}_{\bot
}^{T}>c $ because $\xi <1$ in the used approximation) appears as a
consequence of a genuine variation of the refraction index in the considered
space-time.

Modifications of the photon speed in non-trivial vacua have been previously
reported in the literature (for a review see Refs. \cite{Scharn1}, and \cite
{Pascual}). For example, in the case of the Casimir vacuum\cite{Scharn2},
\cite{Barton} a superluminal light velocity was found, while in the thermal
vacuum an infraluminal velocity was obtained \cite{Pascual},\cite{Barton},%
\cite{Tarrach}. We stress that in both cases the modifications appear as
two-loop effects. In the $S^{1}\times R^{3}$ domain, however, the
modifications that we are obtaining are one-loop correction and therefore
more significant. It is easy to corroborate that for compactification
lengths in agreement with the used approximation, $a<1/m\sim 10^{3}\,fm$,
the transverse velocity (\ref{18}) is about 0.1\% larger than the light
velocity in trivial space-time.

We would like to address a point of possible concern. It has to do with the
fact that albeit ${\rm v}_{\bot }^{T}>c,$ there is no causality violation
here. To understand this, let us recall that the velocity (\ref{18}) is a
low-frequency mode velocity. On the other hand, the velocity of interest for
signal propagation, and hence the relevant one for causality violation
problems, is the high-frequency velocity ${\rm v}_{\bot
}^{T}(q_{0}\rightarrow \infty )$. To determine the difference between the
two, it would be needed to investigate the absorption coefficient $%
\mathop{\rm Im}%
\,n(q_{0})$, with $n(q_{0})$ being the refraction index as a function of the
frequency in the space with $S^{1}\times R^{3}$ topology. However, beyond
any needed calculation, we agree with the analysis of Refs. \cite{causality}
about the lack of causality violations in similar systems. We believe that
in the case under study no (micro-)causality should be violated, because the
events taking place in the $S^{1}\times R^{3}$ space are not constrained by
the null cone of a Minkowskian system, as Lorentz symmetry is explicitly
broken in the present situation.

The low-frequency limit ( $k_{0}=0$, $\left| {\bf k}\right| \rightarrow 0$)
used to obtain the longitudinal-mode mass $M_{\bot }^{L}$ is essential to
study the static properties of the electromagnetic field in this space. The
mass obtained in this limit plays the role of a magnetic mass of the
longitudinal electromagnetic mode\cite{Ferrer}. Below, we show how this
topological mass affects the magnetic response of the system.

In linear response theory, the Maxwell equation of the longitudinal mode
propagating perpendicularly to the ${\cal OZ}$-direction, taken in the
Lorentz gauge $\partial _{\mu }A^{\mu }=0$, is

\begin{equation}
\left[ \Box \delta _{\mu \nu }-(\lambda _{\bot }^{L})^{-2}\delta _{3\mu
}\delta _{3\nu }\right] A_{\nu }=eJ_{\mu },  \label{29}
\end{equation}
where the magnetic length $\lambda _{\bot }^{L}$ is given, as usual, by $%
\lambda _{\bot }^{L}=1/M_{\bot }^{L}$. Considering an external static and
constant current flowing inside the confined space along the ${\cal OZ}$%
-axis, $eJ_{3}(x)=I\,\delta ^{2}({\bf x}_{\bot })$, the induced potential,
which is solution of Eq. (\ref{29}) with periodic boundary conditions in the
${\cal OZ}$-direction, is

\begin{equation}
A_{3}({\bf x}_{\bot })=-\frac{I}{2\pi }K_{0}\left[ \left| {\bf x}_{\bot
}\right| /\lambda _{\bot }^{L}\right]  \label{30}
\end{equation}
where $K_{0}$ denotes a modified Bessel function of the third kind. In the $%
\left| {\bf x}_{\bot }\right| \gg \lambda _{\bot }^{L}$ limit the
corresponding magnetic field is

\begin{equation}
{\bf B}(r)=\frac{I}{2\sqrt{2\pi }}\frac{\exp \left( -r/\lambda _{\bot
}^{L}\right) }{\sqrt{r\lambda _{\bot }^{L}}}\widehat{{\bf \theta }}
\label{31}
\end{equation}
with $r=\left| {\bf x}_{\bot }\right| $ and $\widehat{{\bf \theta }}$
denoting the azimuth-angle unit vector of the cylindrical coordinates taken
with $z$ along the ${\cal OZ}$-axis. From (\ref{31}) we see that an
azimuthal magnetic field will be screened along the radial direction, in a
distance equal to the inverse of the topological mass $M_{\bot }^{L}$. The
smaller the compactification length $a$, the larger the screening in the
transverse direction.

Let us turn our attention now to photons propagating along the ${\cal OZ}$%
-direction ($k_{\perp }=0$). In this case the dispersion relations (\ref{11}%
) and (\ref{12}) can be written respectively as

\begin{equation}
k_{0}^{2}-k_{3}^{2}-\Pi _{11}=0  \label{32}
\end{equation}

\begin{equation}
k_{0}^{2}-k_{3}^{2}-\frac{k_{0}^{2}-k_{3}^{2}}{k_{0}^{2}}\Pi _{33}=0
\label{32a}
\end{equation}

Assuming $ak_{0}\ll am\ll 1$ in (\ref{15}), the components of the
polarization operator appearing in (\ref{32}) and (\ref{32a}) become

\begin{equation}
\Pi _{11}(k_{\bot }=0,k_{3},k_{0}\sim 0)\simeq \frac{e^{2}}{a^{2}}\left[
\frac{1}{9}+{\cal O}(\xi ^{2})\right] +{\cal O}(k_{0}^{2})  \label{33}
\end{equation}
\begin{equation}
\Pi _{33}(k_{\bot }=0,k_{3},k_{0}\sim 0)\simeq \frac{e^{2}}{a^{2}}\frac{%
k_{0}^{2}}{k_{3}^{2}}\left[ \frac{-1}{9}+{\cal O}(\xi ^{2})\right] +{\cal O}%
(k_{0}^{4})  \label{33a}
\end{equation}

From (\ref{32}), (\ref{32a}), (\ref{33}) and (\ref{33a}) we can
straightforwardly find the low-frequency limit of the photon velocities for
transverse and longitudinal modes propagating along the ${\cal OZ}$-direction

\begin{equation}
{\rm v}_{\Vert }^{T}\simeq 1-\left[ (M_{\Vert }^{T})^{2}/2k_{3}^{2}\right] ,
\label{34}
\end{equation}
\begin{equation}
{\rm v}_{\Vert }^{L}\simeq 1-\left[ (M_{\Vert }^{L})^{2}/2k_{3}^{2}\right] ,
\label{34a}
\end{equation}
where both transverse and longitudinal mode masses coincide and are given by
$M_{\Vert }^{T}=M_{\Vert }^{L}=e/3a$. We stress that in this case both
velocities are smaller than the light velocity in trivial Minkowski space $%
c, $ and that the modification is due, as in (\ref{18a}), to the appearance
of a topological photon mass for each mode.

One can investigate, as above, the implications of the masses $M_{\Vert
}^{T} $ and $M_{\Vert }^{L}$ for the magnetic response of the system. A mass
for a mode propagating along ${\cal OZ}$ yields a magnetic screening in that
direction. However, since the effective penetration lengths are $\lambda
_{\Vert }^{T}=\lambda _{\Vert }^{L}=3a/e>a,$ the corresponding magnetic
fields would decay in a distance larger than the compactification length,
meaning that the screening effect is unobservable.

In summary, we have shown in this paper that in the $S^{1}\times R^{3}$
space-time the photon propagates through transverse and longitudinal modes.
In addition, each mode moves anisotropically along the two identified
spatial directions (parallel and perpendicular to ${\cal OZ)}$. This effect
can be interpreted as a topological birefringence. Only the transverse mode
moving perpendicularly to the compactified direction, remains massless. It
propagates with a superluminal group velocity that increases logarithmically
with the compactification radius $a$. The remaining modes acquire
topological masses inversely depending on $a$. As a consequence, at low
energies ($E<M_{i}$, $M_{i}=\{M_{\bot }^{L},M_{\Vert }^{L},M_{\Vert }^{T}\}$%
) the massive photons decouple and only the transverse mode propagating with
velocity ${\rm v}_{\bot }^{T}$ will be active. This implies that at low
energies the topological mass generation acts as a natural confining
mechanism that prevents photon propagation along the compactified direction.
Therefore, even though the original $S^{1}\times R^{3}$ space-time is a
four-dimensional domain, it reduces at low energies to an effective $(2+1)-$%
dimensional manifold for the fields $A_{\mu }$. These results can be of
interest for extra-dimension and brane theories, where it is necessary to
confine at low energies the gauge fields to the branes.

Another important result is the existence of a topologically induced
directional superconductivity, with Meissner effect taking place for
magnetic fields contained in the transverse $(2+1)-$dimensional space. This
result could find applications in high-T$_{C}$ superconductivity where, as
it is well known, the system is confined to a quasi-two-dimensional space.

{\bf Acknowledgments }

It is a pleasure for two of the authors (EJF and VI) to express their
gratitude to the Institute for Spatial Studies of Catalonia for the warm
hospitality extended to them during the time this work was completed, and in
particular to Dr. E. Elizalde for facilitating this collaboration. This work
has been supported in part by US-Spain Science and Technology CCCT-6-151
grant, by NSF grant PHY-0070986 (EJF and VI), and by NSF POWRE grant
PHY-9973708 (VI).

\end{document}